\documentclass[twocolumn]{aastex631}

\usepackage{graphicx}	
\usepackage{float}      
\usepackage{amsmath}	
\usepackage{amsfonts}
\usepackage{cancel}
\usepackage{amsfonts}
\usepackage{placeins}

\usepackage{hyperref}
\hypersetup{
    colorlinks=true,     
    linkcolor=blue,      
    citecolor=blue,      
    urlcolor=blue        
}

\usepackage{fancyhdr}
\newcommand{\insitu}{\textit{in-situ }}
\newcommand{\exsitu}{\textit{ex-situ }}

\begin{document}

\title{Tracing the relic nature of compact galaxies through their globular cluster systems}

\correspondingauthor{Micheli T. Moura}
\email{micheli.t.moura@gmail.com}

\author[0000-0002-0786-7307]{Micheli T. Moura}
\affiliation{Departamento de Astronomia, Instituto de Astronomia, Geofísica e Ciências Atmosféricas, Universidade de São Paulo, Rua do Matão 1226, Cidade Universitária, 05508-900, São Paulo, SP, Brazil}
\affiliation{Instituto de Física, Universidade Federal do Rio Grande do Sul, Av. Bento Gonçalves 9500, Porto Alegre, R.S. 90040-060, Brazil}

\author[0000-0003-3220-0165] {Ana L. Chies-Santos}
\affiliation{Instituto de Física, Universidade Federal do Rio Grande do Sul, Av. Bento Gonçalves 9500, Porto Alegre, R.S. 90040-060, Brazil}
\author[0000-0001-6838-431X]{Cristina Furlanetto}
\affiliation{Instituto de Física, Universidade Federal do Rio Grande do Sul, Av. Bento Gonçalves 9500, Porto Alegre, R.S. 90040-060, Brazil}
\author[0000-0002-5970-2563]{Yingtian Chen}
\affiliation{Department of Astronomy, University of Michigan, Ann Arbor, MI 48109, USA}
\author[0000-0001-9852-9954]{Oleg Y. Gnedin}
\affiliation{Department of Astronomy, University of Michigan, Ann Arbor, MI 48109, USA}
\author[0000-0002-4694-2250]{Michael A. Beasley}
\affiliation{Instituto de Astrofísica de Canarias, Calle Vía Láctea, E-38206 La Laguna, Spain}
\affiliation{Departamento de Astrofísica, Universidad de La Laguna, E-38206 La Laguna, Spain}
\affiliation{Centre for Astrophysics and Supercomputing, Swinburne University, John Street, Hawthorn VIC 3122, Australia}
\author[0000-0002-6411-220X]{Anna Ferré-Mateu}
\affiliation{Instituto de Astrofísica de Canarias, Calle Vía Láctea, E-38206 La Laguna, Spain}
\affiliation{Departamento de Astrofísica, Universidad de La Laguna, E-38206 La Laguna, Spain}
\affiliation{Centre for Astrophysics and Supercomputing, Swinburne University, John Street, Hawthorn VIC 3122, Australia}
\author[0000-0002-8005-0870]{Ling Zhu}
\affiliation{Shanghai Astronomical Observatory, Chinese Academy of Sciences, 80 Nandan Road, Shanghai 200030, China}
\author[0000-0002-0786-7307]{Juan Pablo Caso}
\affiliation{Facultad de Ciencias Astronomicas y Geofísícas de la Universidad Nacional de La Plata, and Instituto de Astrofísica de La Plata (CCT La Plata – CONICET, UNLP), Paseo del Bosque S/N, B1900FWA La Plata, Argentina}

\begin{abstract}
We investigate the synthetic model of globular cluster (GC) systems of 17 compact massive galaxies (CMGs) from the Illustris TNG100 simulation to explore their connection with massive relic galaxies, systems that have undergone little structural evolution across cosmic time. The co-evolution of the GC systems and their host galaxies is based on a GC formation and evolution model that assigns clusters to stellar particles according to age and local conditions, providing positional, kinematic, and chemical information for individual GCs. By combining stellar assembly histories, effective radius evolution, and GC properties such as \insitu vs. \exsitu origin, metallicity, and spatial distribution, we identify consistent signatures of early formation and late-time accretion. We find that the GC mass fraction traces the host assembly history more robustly than the GC number fraction, as massive clusters better preserve the imprint of the early accretion history. Three CMGs from TNG100 emerge as strong massive relic analogs, exhibiting high \insitu GC fractions, narrow metallicity distributions, and compact spatial distributions. A tight correlation between the host stripped fraction and the extent of the \exsitu GC population further reveals the possibility to consider GC spatial profiles as signature to identify tidal stripping processes. These results indicate that the combined analysis of GC populations and host stellar assembly offers a robust diagnostic for identifying massive relic galaxies and constraining their evolutionary histories.

\end{abstract}

\section{Introduction} 
\label{sec:intro}

While most massive galaxies in the local Universe are extended systems with old stellar populations, a surprising minority retain the compact structures ($\rm{R_e<2}\,kpc$) characteristic of high-redshift progenitors (e.g.,\,\citealt{2005Daddi,2009Damjanov}). According to the widely adopted two-phase formation scenario (e.g.,\,\citealt{2009Naab,2010OserOstriker}), massive galaxies first form rapidly through dissipative processes at high redshift, giving rise to compact and dense systems called \textit{red nuggets}. In a subsequent phase, they undergo gradual size and mass growth primarily through dry minor mergers, leading to the extended early-type galaxy (ETG) population observed today \citep{vanderWel2014,2021Davison}. However, a small fraction of these early-formed systems appears to have avoided significant mass or structural evolution, remaining compact until the present day.

These galaxies, often referred to as relics (e.g.,\,\citealt{2009Trujillo,Ferre-Mateu2017,2021Spiniello,2023Lisiecki,2023Grebol}), are considered the most pristine descendants of those red nuggets and offer a unique window into the early phases of massive galaxy formation. Among their properties, relic galaxies are massive and compact ETGs which typically show no clear morphological signs of recent mergers or interactions \citep{2014Trujillo,2018ferre-mateu,2024spiniello}. Their kinematic properties, including cold stellar orbits and distinct inner halo dynamics compared to normal ETGs, support the scenario that these systems have evolved passively since their early formation epoch \citep{2024Moura, 2025Zhu}.

Results from the INSPIRE project \citep{2021Spiniello,2024spiniello} have provided a quantitative framework to characterize the evolutionary state of relic candidates through the Degree of Relicness (DoR, \citealt{Ferre-Mateu2017}). The DoR is defined as a dimensionless value ranging from 1 to 0, where 1 represents the most extreme case of a relic galaxy (e.g., NGC 1277 in the local Universe is the prime example of an extreme relic, with a DoR of almost 1), while 0 represents a galaxy still forming stars \citep{2024spiniello}.


In typical massive ETGs, GC systems usually display a bimodal optical color, often displayed as metallicity differences (e.g. \citealt{2012Chies-santos}, \citealt{blakeslee12}, \citealt{2016Cho}, \citealt{2020Fahrion}), with metal-rich (red) GCs believed to form mostly \insitu during the early dissipative collapse and metal-poor (blue) GCs largely accreted through subsequent mergers \citep{2006Brodiestrader,2018forbes}. In this context, GC systems provide an effective means of distinguishing massive relic from non-relic galaxies. A predominance of old, metal-rich GCs is expected for massive relics, reflecting their early, \insitu-dominated formation and limited subsequent growth, whereas systems with a significant fraction of metal-poor GCs are indicative of extended accretion histories \citep{2020Beasley}.

Most of the GCs are generally found to be older than 10 Gyrs \citep{strader05,2012Chies-santos, usher19}, the variety of some color distributions can also be partially attributed to age differences \citep{2019Lee}. In this context, massive relic galaxies are expected to be dominated by old red GCs, as their formation is generally thought to be confined to an early \insitu phase with minimal later accretion.

To date, observational studies of GC systems in confirmed massive relic galaxies remains limited. The GC system of NGC\,1277, our \textit{eximium exemplum} massive relic galaxy, has been shown to be mainly dominated by red GCs (considering the $g$-$z$ color) with a red fraction exceeding 80\%. This is consistent with expectations for a red-nugget galaxy that remained mainly unaltered during the second phase of its evolution \citep{2018Beasley}, considering a scenario where the accreted component is generally more metal-poor, therefore comprising also blue GCs. Beyond NGC\,1277, studies have focused on broader samples of compact massive galaxies, particularly adopting the \citet{2017Yildrim} sample. Two works\,--\,\citet{2021Kanglee} and \citet{2021Alamo}\,--\,analyzed overlapping selections from this sample using archival \textit{Hubble Space Telescope} (HST) data, focusing on the properties of the GCs population of these compact ETGs. \citet{2021Kanglee} studied 12 of these massive compact ellipticals (CMGs), finding a wide range in red GC fractions ($f_{\mathrm{RGC}} = 0.2$--0.7, mean $\sim0.48$). They reported an environmental dependence, where CMGs in clusters had higher $f_{\mathrm{RGC}}$ ($\sim0.60$) than group/field counterparts ($\sim0.40$), suggesting that dense environments favours red GCs. Meanwhile, \citet{2021Alamo} examined 15 massive compact ETGs (13 overlapping with \citealt{2017Yildrim} sample), revealing some trends, as lower specific GC frequencies ($S_N < 2.5$, median $S_N = 1$) compared to normal ellipticals, supporting formation via rapid early collapse, and an anticorrelation between host galaxy spin ($\lambda_R$) and GC color dispersion ($I$--$H$), implying that mergers broaden the GC color distribution.

However, recent kinematical modelling analyses through orbital decomposition by \citet{2025Zhu} caution that not all galaxies in the \citet{2017Yildrim} sample are confirmed massive relics, since some show merger signatures. This highlights the need to combine multiwavelength GC studies with simulations to robustly add an additional signature on the properties of GCs in relic systems and reconstruct their evolution in connection with their hosts.

In this work, we select a sample of compact massive early-type galaxies (CMGs) from the IllustrisTNG (e.g.,\,\citealt{2018Nelson,2019Pillepich}) simulation (TNG100) and apply an analytical framework to investigate the assembly history of their associated globular cluster systems. We aim to characterize the chemical and spatial properties of GC populations and investigate how they relate to the assembly histories of their compact massive hosts, focusing on the roles of accretion and tidal stripping. This paper is organized to introduce the simulation sample selection and the employed methods of GCs formation and evolution in Section \ref{sec:methods}. Analysis, results and discussion are presented in Sections \ref{sec:analysis}, \ref{sec:results}, and \ref{sec:discussion} respectively. Throughout the paper, we assume the \citet{plank2016} parameters to $\Lambda$CDM cosmological model with $H_{0} = 67.74\,\rm{km}\,\rm{s}^{-1} \rm{Mpc}^{-1}, \Omega_{m} = 0.30$, and $\Omega_{\Lambda} = 0.69$.

\section{Methods and sample selection}\label{sec:methods}

\subsection{TNG100 simulation}
\label{sample}

To investigate the evolutionary assembly of CMGs, we use the IllustrisTNG suite of cosmological magnetohydrodynamical simulations (e.g.,\,\citealt{2018Pillepich,2018Springel,2018Nelson,2019Nelson}), with the moving-mesh code \textsc{arepo} \citep{2010Springel}. Specifically, we adopt data from the TNG100-1 run, which offers a good compromise between volume and resolution, covering a box of 110.7 co-moving Mpc with baryonic mass resolution of $1.4 \times 10^6 \,\rm{M}_\odot$ and dark matter particles of $7.5 \times 10^6\,\rm{M}_\odot$.

Galaxies and halos are identified through a combination of the friends-of-friends (FoF) algorithm \citep{1982HuchraGeller} and \textsc{subfind} \citep{2001Springel,2009Dolag}, which locates gravitationally bound substructures. Subhalos are linked across time using merger trees \citep{2015Rodriguez}, allowing us to track the assembly history and classify stellar particles as \textit{in-situ} or \textit{ex-situ} based on their origin. Specifically, a stellar particle is classified as \textit{in-situ} if it originated within a galaxy that belongs to the main progenitor branch of its current host galaxy. If it formed in a different galaxy that does not lie along this main evolutionary path, it is considered \textit{ex-situ} \citep{2016Rodriguez}.

Halo properties are defined within $R_{200}$, the radius enclosing a mean density 200 times the critical density at each redshift. The resolution and volume of TNG100 together provide the necessary detail to study massive galaxy evolution across cosmic time with robust statistics. Stellar masses are computed as the total mass of bound star particles within a given aperture, while galaxy sizes are typically defined based on the three-dimensional half-mass radius of the stellar distribution. The radius, in this context, corresponds to the half-mass radius, which encloses half of the total stellar mass of the galaxy. 

\begin{figure}
\centering
\includegraphics[width=\columnwidth]{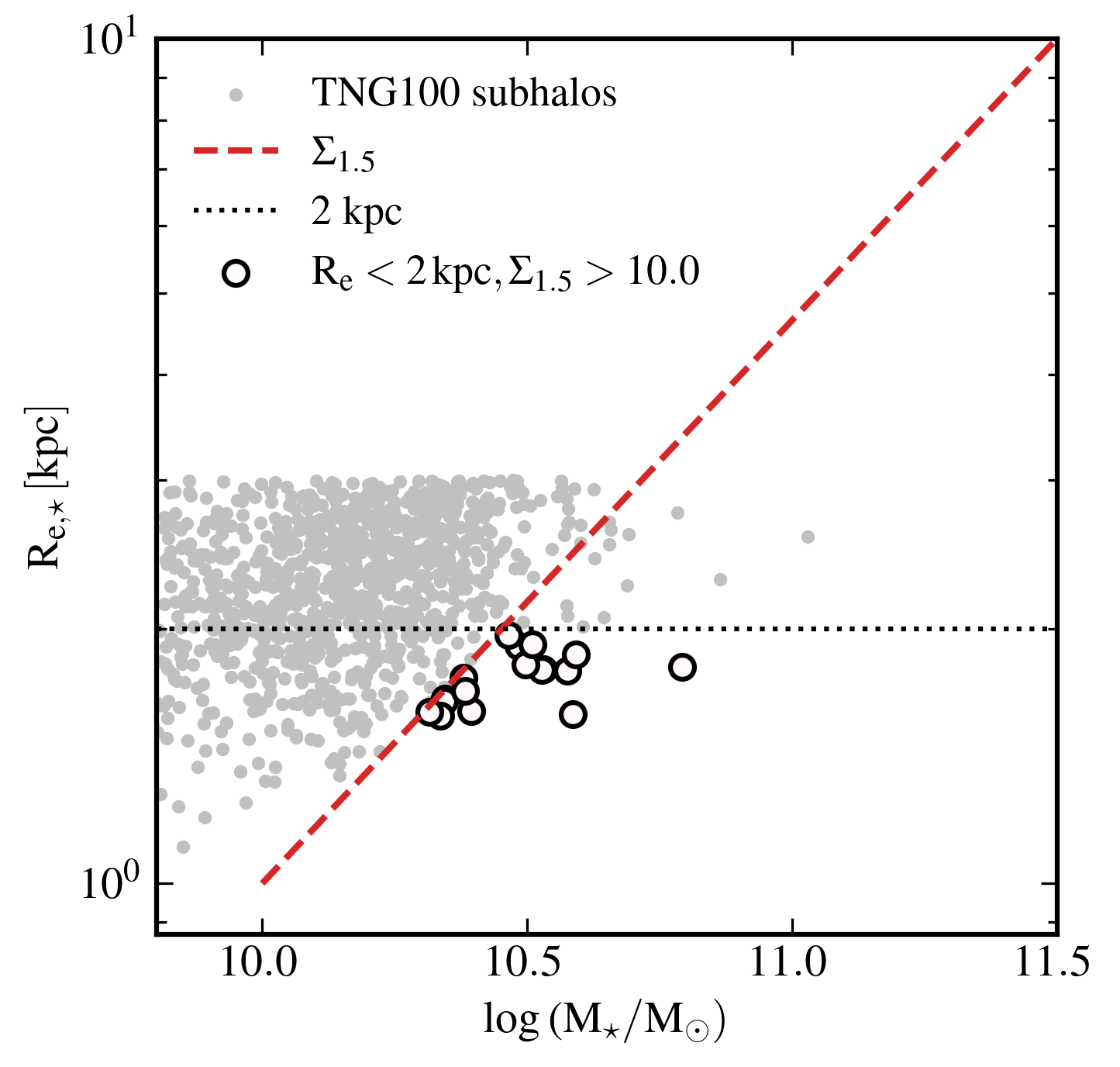}
\caption{Stellar mass--size relation for the selected sample of CMGs at $z=0$. A sample of TNG100 subhalos are shown as light gray circles for reference, while selected compact subhalos are represented by open circles. Dotted black line indicates the threshold of 2 kpc, while the dashed red line follows the surface density parameter for the selected sample ($\rm{R}_e<2\,\rm{kpc}, log\,\Sigma_{1.5}>10.0\,dex$).}
\label{fig:mass-size_selection}
\end{figure}

To select our simulated sample, we selected all subhalos with stellar masses above $10^{10}\,\rm{M_\odot}$ and sizes $R_e\,<2\,\rm{kpc}$ to match with observations (e.g.,\,\citealt{2023GrebolTomas}) and previous searches in simulations \citep{2025Moura}. The mass-–size relation of TNG galaxies has been extensively compared with SDSS observations, showing good agreement in general in our range of masses and sizes \citep{2018Genel}. The final selection of subhalos can be seen in the mass--size plane in Fig.~\ref{fig:mass-size_selection}.

We use \citet{2013Barro} surface density as a third constraint to select all the compact subhalos among this first selection \citep{2023GrebolTomas,2025Zhu}. The compactness criterion considering the surface density is defined as $\Sigma_{1.5} = \rm{M\,[M_\odot]/(R_e\,[kpc])^{1.5}}$, where the stellar mass within the effective radius is considered. We choose a less strict threshold of $\Sigma_{1.5} > 10.0$\,dex, instead of the 10.3\,dex adopted in \citealt{2013Barro, 2023Grebol,2025Moura} in order to enlarge our sample size. This adjustment is not motivated by a systematic size offset between simulations and observations, but by sample statistics, since only three subhalos satisfy the original 10.3 dex criterion in our sample. Lowering the threshold to 10.0 dex increases the sample size while still selecting highly compact systems, as $\Sigma_{1.5}$ is a continuous compactness parameter and values above 10 dex remain within the regime typically associated with compact galaxies. A total of 21 subhalos meet the selection criteria; however, 4 were excluded due to irregular morphology noticed by visual inspection. The final sample therefore consists of 17 subhalos, whose properties are presented in Table~\ref{tab:gc_fractions} in Appendix~\ref{appendixA}.


\subsection{The model for globular cluster formation and evolution}
\label{GC_model}
We adopt the model for GC formation and evolution developed in \citet{2010Muratov}, \citet{2019Lee}, and \citet{2019Choski}, which uses the halo merger tree as input to trigger the GC formation when the specific mass accretion rate exceeds a predefined threshold. Based on a sequence of scaling relations, the model analytically computes the mass and metallicity of newly-formed GCs \citep{2023Chen,2024Chen}. It also accounts for mass loss due to stellar evolution and tidal disruption. The version of the model used in \citet{2022Chen} improves upon earlier implementations by incorporating spatial and kinematic information for GCs, using stellar and dark matter particles as tracers from the selected simulation.

The implementation of the model follows determined steps: cluster formation, cluster sampling and its particle assignment, and cluster evolution. GC formation is triggered by rapid mass growth of the host galaxy, such as during major mergers or periods of intense accretion of material (gas, stars). This is quantified by the specific mass accretion rate, defined as the fractional change in galaxy mass between two consecutive simulation snapshots. When this rate exceeds a predefined threshold, a GC formation event is triggered.

The total mass of the newly formed GC population is then calculated using the linear relation between cluster mass and gas mass \citep{2005Kravtsov}. The simulated metallicity of the clusters is assigned following \citet{2018Choksi}. In this approach, it is drawn from the simulated metallicity of the host galaxy’s interstellar medium, which is modeled as a double power-law function of stellar mass and redshift. This prescription is observationally calibrated and accounts for the presence of more metal-rich populations in giant elliptical galaxies.

After formation, the number of individual GCs is sampled from a \citet{1976Schecter} initial cluster mass function (ICMF). These clusters are then assigned to simulation particles. In hydrodynamic simulations like TNG100, young stellar particles (age $<10$~Myr) are selected to host the newly formed clusters, reflecting the typical timescale of cluster formation. The cluster evolution is then tracked by following the positions and velocities of these particles, allowing the model to trace the dynamical history of individual GCs and compute their evolving properties analytically. This approach reduces dependency on the specific baryonic physics implemented in the hydrodynamic simulation.

The model also accounts for GC mass loss due to stellar evolution and dynamical disruption. Disruption is modeled using the formalism of \citet{2023Gieles}, where the rate of mass loss is described by a multivariate power-law function of both the initial and current cluster mass, and depends on the local tidal field along each GC’s orbit. 

The model has accurately reproduced properties of observed GC systems, such as the distributions of mass, metallicity, galactocentric distance, velocity dispersion, and orbital anisotropy, for both Milky Way-mass galaxies \citep{2022Chen} and dwarf galaxies \citep{2023Chen}. When applied to galaxies with similar masses but different merger histories, the model also captured the higher metallicity observed in M31-like galaxies that recently experienced a major merger, relative to the more quiescent Milky Way analogs \citep{2024Chen}. Extending to a broader range of galaxy masses, it further recovered global scaling relations, including the effective radius–galaxy mass relation and the nearly linear correlation between GC system mass and host galaxy mass \citep{2024Chen}.

Based on the accuracy of the model applied to Milky Way and dwarf-mass galaxies, we now extend it to CMGs in the TNG100 simulation, selecting only survived GCs at $z=0$, and with masses greater than $10^{5}\,\rm{M_\odot}$. This method enables us, for the first time, to investigate the evolution of GC systems of compact massive galaxies, providing a test for the two-phase formation scenario and offering a new piece to the puzzle of massive system evolution.
\newline

%
\begin{figure}
\centering
\includegraphics[width=\columnwidth]{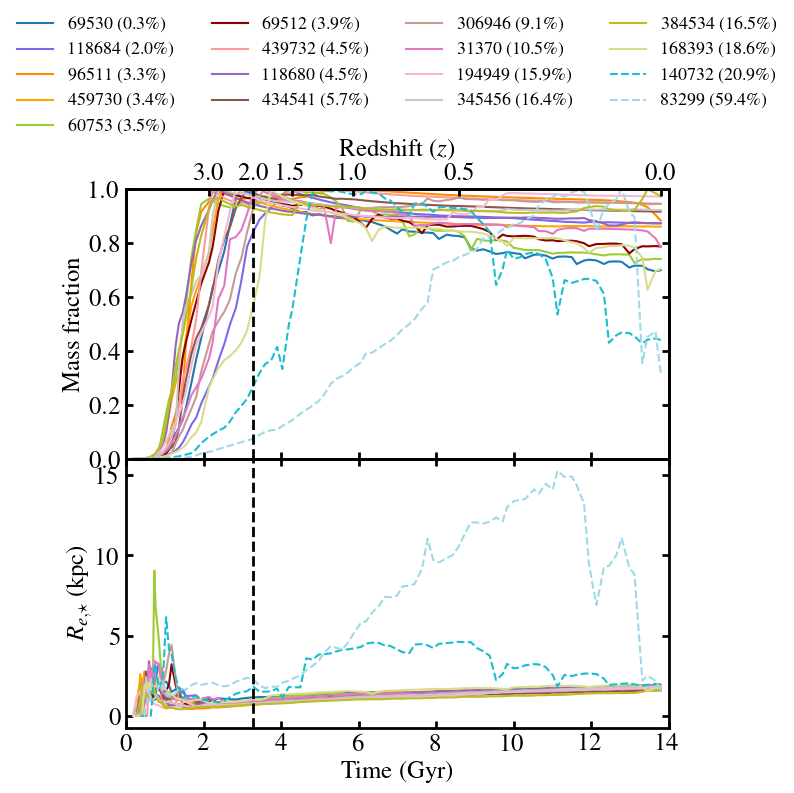}
\caption{Stellar mass assembly histories of the compact massive host galaxies as a function of cosmic time. The stellar mass enclosed within the effective radius, $R_{e,\star}$, is normalized to its value at $z=0$. Each host galaxy is shown in colored lines, where solid lines correspond to systems with satellite accretion fractions below $20\%$, while dashed lines indicate galaxies with more extended accretion histories ($>20\%$). The vertical line marks $z = 2$, commonly adopted as the onset of the second phase of massive galaxy evolution. Overall, $58.8\%$ of the hosts have accretion fractions below $10\%$, while $88.2\%$ have \textit{ex-situ} stellar mass fractions below $20\%$.}
\label{fig:host_assembly_hist}
\end{figure}
\begin{figure*}
\centering
\includegraphics[width=\linewidth]{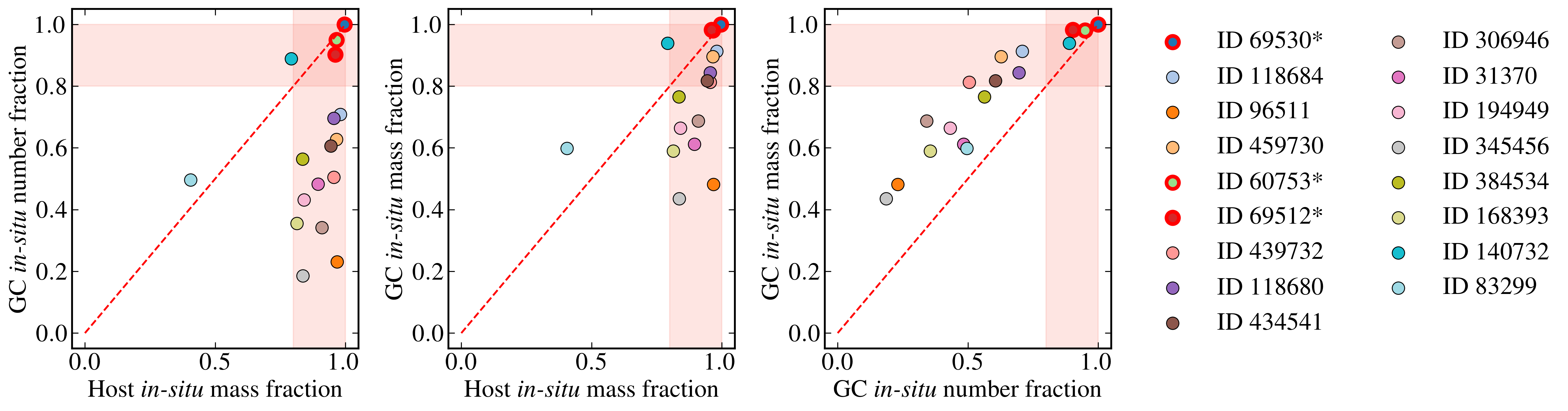}
\caption{GC \insitu fraction \textit{vs}. host \insitu mass fraction. The diagonal dashed line indicates the 1:1 relation. The left panel shows the GC \insitu ($N_{gc}$) number fraction as a function of host \insitu mass fraction, the center panel shows the GC \insitu mass fraction as a function of host \insitu mass fraction, and the right panel shows the GC number \textit{vs.} the GC mass fraction. We are considering just GCs with masses higher than $10^5\,\rm{M_\odot}$ to compute the fractions. The shaded region indicates where the \insitu fractions are $>80\%$ for \textit{x} and \textit{y}-axis. Similarly, the `*' symbol in the legend indicates the host IDs where the \insitu fraction (for hosts and GCs) is $>90\%$ among the three panels.}
\label{fig:similarity2}
\end{figure*}

\section{Analysis} \label{sec:analysis}

\subsection{Assembly histories of hosts and their GCs}\label{host_assembly}

Based on the selection criteria related to mass, size, and compactness described in Sect.\,\ref{sample}, we present the stellar assembly history of the 17 selected galaxies in Fig.~\ref{fig:host_assembly_hist}. The upper panel shows the stellar mass within $1\,\rm{R_e}$ at each redshift normalized by the galaxy's maximum stellar mass ($\rm{M_\star / M_{\star(max)}}$). The lower panel displays the evolution of the effective radius over time for each galaxy. In both panels, galaxies are color-coded according to their \textit{ex-situ} stellar mass fraction computed from the stellar assembly catalogue for TNG100 \citep{2016Rodrigues-gomez}. We will assume the notation of \textit{`host'} to refer to these galaxies from now on.

This assembly information is important because it reveals the individual star formation histories (SFHs) of each galaxy until $z=0$. While all selected objects meet the CMGs criteria (considering mass, size, and compactness), not all CMGs exhibit early or fast stellar assembly (e.g., IDs 140732 and 83299). An extended SFH suggests that the galaxy may have experienced mergers during the second phase of its evolution (for a detailed discussion, see \citealt{2023Grebol}), potentially followed by tidal stripping or other environmental effects, which turn them compact at $z=0$. As a result, the 17 host galaxies exhibit varying \exsitu stellar mass fractions, with some showing extended SFHs and signs of tidal stripping (i.e.,\,IDs 140732 and 83299).

Overall, 15 out of 17 systems ($\approx88\%$) show \exsitu stellar mass fractions below 20\%, while the remaining two ($\approx12\%$) present higher \exsitu fractions ($>20\%$) and clear signatures of stellar tidal stripping (Fig.~\ref{fig:host_assembly_hist}), in Sect.\,\ref{stripping}, we will discuss the stellar stripping of these systems in more detail. \citet{2018Beasley} estimated $\approx12\%$ of \exsitu fraction for NGC\,1277, indicating that some of our selected hosts share similar accretion histories with this well-studied relic galaxy. Cosmological simulations predict that the fraction of accreted stellar mass increases with galaxy mass. For galaxies with stellar masses $M_\star \sim 10^{10.5} - 10^{11}\, M_\odot$, typical \textit{ex-situ} fractions range from $\sim 10\%$ to $\sim 50\%$ (e.g.,\,\citet{2016Rodrigues-gomez}). As reference, observational estimates based on stellar population and dynamical modeling also suggest comparable values, typically $\sim20–60\%~f_{\rm exsitu}$ for massive ETGs (e.g.,\,\citealt{2021Davison,2024Angeloudi}).

For all the selected host CMGs, we investigated the assembly of their GC systems, distinguishing the \textit{in-situ} and \textit{ex-situ} components. Given that relic galaxies are expected to contain negligible \textit{ex-situ} material, we examined the relation between the host \textit{in-situ} stellar mass fraction and the corresponding \textit{in-situ} GC fractions, both in number and mass. From this relation, we identified the most extreme cases of relic-like galaxies within our CMG sample, which will serve as benchmarks for exploring parameters as GC metallicities and kinematics among the early assembly of massive galaxies.

The model described in Sect.\,\ref{GC_model} was applied to the 17 host halos to identify the associated GCs. In this framework, GCs that formed within the galaxy's main progenitor branch are classified as \textit{in-situ}, while those that originated in satellite galaxies are considered \textit{ex-situ}. The \textit{ex-situ} clusters are later brought into the central galaxy through accretion and mergers \citep{2022Chen}.

Since our goal is to investigate the connection between the host galaxy and the properties of its GCs, it is essential to link the accretion and evolutionary histories of both components. To this end, Fig.\,\ref{fig:similarity2} presents the \textit{in-situ} fractions for both the host galaxies and their GCs, enabling a direct comparison of their accretion histories. 

Here we explore two approaches to evaluate the assembly of GCs in connection with the assembly of their host galaxies. The first approach is based on the number of GCs ($N_{\mathrm{GC}}$), where the \insitu fraction is defined as: 
$f_{\mathrm{N_{GC}}} = N_{{\textit{in-situ}}}/{N_{\rm{total}}}$. Alternatively, the \insitu fraction can be computed based on the total mass of the GCs: $f_{\mathrm{mass}} = {M_{\mathrm{\textit{in-situ}}}}/{M_{\rm{total}}}$. 
We restrict our analysis to GCs with masses above $10^5\,\mathrm{M_\odot}$, approximately the turnover of the observed GC mass function. Including lower-mass clusters would significantly increase the total number of model GCs, as the prescription yields a large low-mass population, thereby biasing number-based statistics without materially affecting mass-weighted trends. We add in Appendix~\ref{appendixA} a comparison between the total number of clusters $N_{GC}$ and the specific frequency $T_N$ (defined as the number of GCs normalized by the host galaxy stellar mass), using data from the ACSVCS \citep{2008Peng} as a reference. The host CMGs lie within the observed range and follow the median relation, indicating that the number of GCs in the simulations is consistent with that observed in galaxies of similar stellar mass.

As shown in Fig.~\ref{fig:similarity2} (left and center panel), the most extreme cases, those with high \insitu GC fraction (i.e $>80-90\%$), whether calculated by number or by mass, may differ depending on which definition of fraction \insitu is adopted.

We found that the average deviation from the 1:1 line (right panel) is larger for the GC number fraction (0.319) than for the GC mass fraction (0.162). The average deviation was computed as the mean absolute difference between the GC and host \textit{in-situ} fractions, i.e., $\langle | f_{\mathrm{GC}} - f_{\mathrm{host}} | \rangle$, using both number- and mass-based GC fractions. This result indicates that the GC mass fraction exhibits a stronger correlation with the corresponding host galaxy fractions, suggesting it as a more reliable tracer of the host assembly history in this context. 

It is also possible that this offset arises because \textit{in-situ} and \textit{ex-situ} clusters have different typical masses. In our sample, the mean GC mass is higher for the \textit{in-situ} population
($\mathrm{mean}\ \log M_{\mathrm{GC}} \approx 5.78$) than for the \textit{ex-situ} population ($\mathrm{mean}\ \log M_{\mathrm{GC}} \approx 5.43$), corresponding to a factor of
$\approx 2$ in mass. This likely reflects both the formation of \insitu clusters in the deeper potential wells of the main progenitor and the preferential survival of more massive clusters in the dense environments of these compact galaxies, several of which experienced significant tidal stripping.

Galaxies located in the extreme regions of the parameter space ($x > 0.8$ and $y > 0.8$) include three systems in the GC number versus host fraction plane (IDs 60753, 69512, 69530), eight in the GC mass versus host fraction plane (IDs 60753, 69512, 69530, 118680, 118684, 434541, 439732, 459730), and four in the GC mass versus number fraction plane (IDs 60753, 69512, 69530, 140732). The galaxies 60753, 69512, and 69530 are common to all three correlations, representing the most GC \textit{in-situ}-dominated cases. When considering the more restrictive threshold ($x > 0.9$ and $y > 0.9$), these same three galaxies remain consistently extreme across all panels, further reinforcing their classification as highly \textit{in-situ}-dominated systems. Therefore these systems represent the best cases for our relic galaxies candidates, considering both, the GC and host assembly.

Although we focus on these three galaxies as the convergent cases with high \textit{in-situ} fractions across all panels, it is worth noting that the subhalos 140732 and 118684 also show a strong consistency with their host assembly, exhibiting GC \textit{in-situ} mass fractions above 80\% and 90\%, respectively. We will include these two subhalos in our discussion, as they would also be considered strong relic galaxy candidates from an observational perspective. In our analysis we consider the pathways of five MCGs (IDs 60753, 69512, 69530, 140732, 118684) exhibiting the strongest correspondence between the GC properties and the host galaxy assembly history, as shown in Fig.\,\ref{fig:similarity2}, where 3 of them (60753, 69512, 69530) are the most ideal cases of relic galaxies candidates.

\begin{figure*}
\centering
\includegraphics[width=\linewidth]{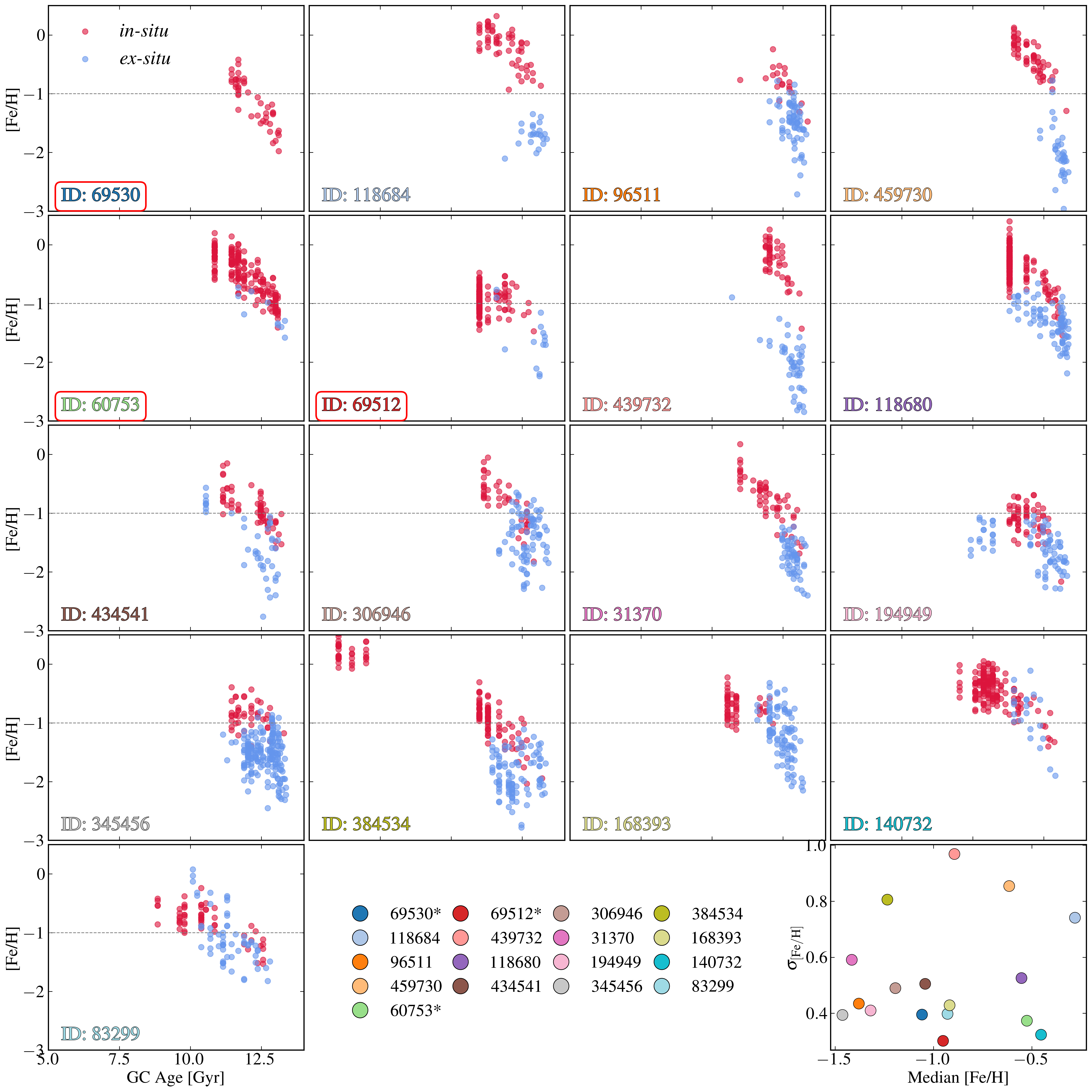}
\caption{[Fe/H] distribution over ages for all GCs in each host galaxy. GCs \insitu are displayed in red, and \exsitu GCs are displayed in blue in all the frames. The horizontal gray indicates the [Fe/H]=$-1$, for reference. The most GC \insitu-dominated hosts are highlighted with `*' symbol. The last row and column display the dispersion $\sigma$[Fe/H] for each host galaxy. The IDs 69530, 60753, 69512 are highlighted with red boxes in the legend frames}
\label{fig:age-met_compacts}
\end{figure*}

\begin{figure*}
\centering
\includegraphics[width=\linewidth]{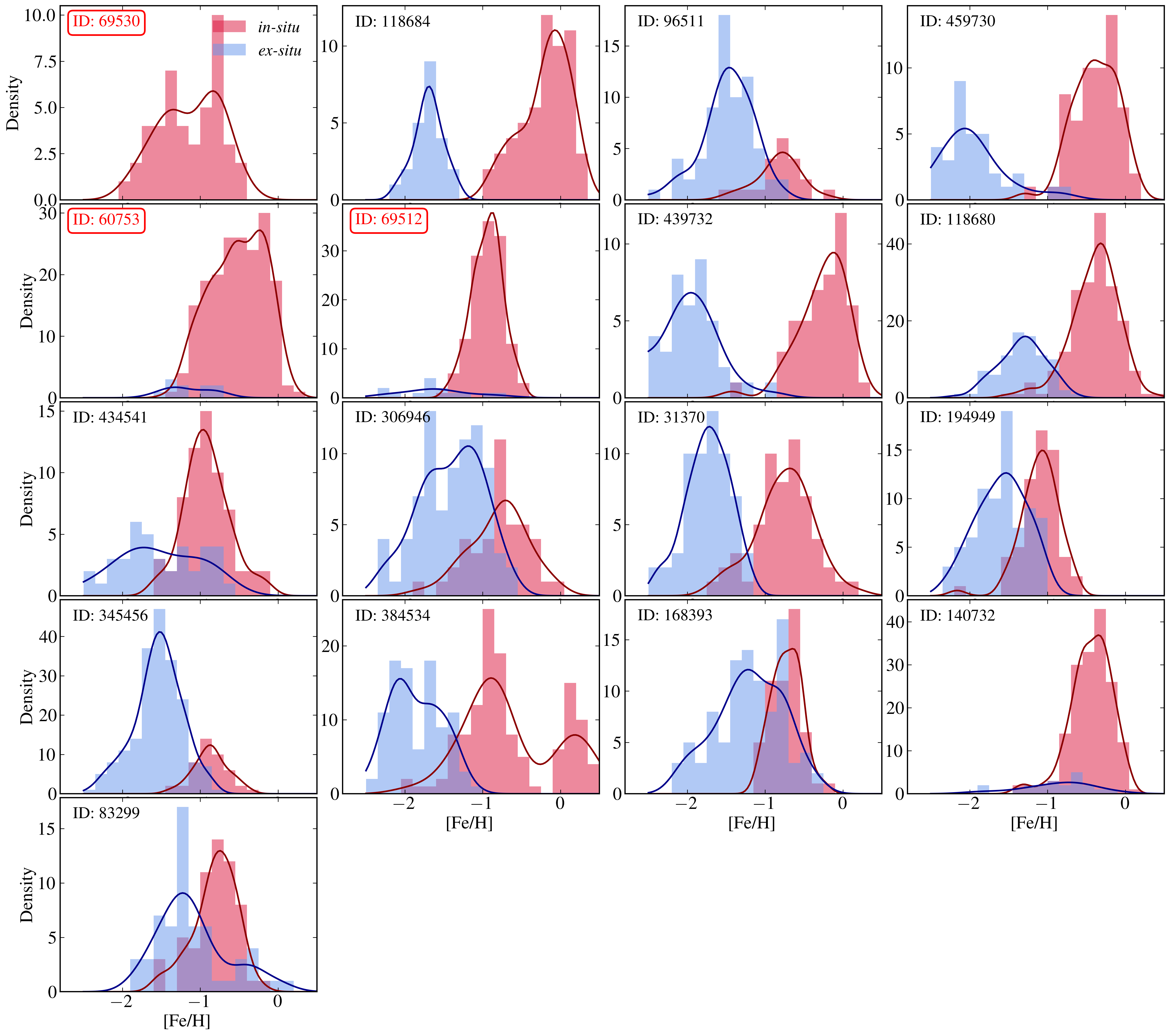}
\caption{Histograms of GC metallicity ([Fe/H]) for each host galaxy in the sample, showing the number of GCs (N$_{\text{GC}}$) as a function of [Fe/H]. The distributions are color-coded in red and blue to represent \insitu and \exsitu populations, respectively. All the frames are ordered to follow the same sequence as in Fig.~\ref{fig:age-met_compacts}.} 
\label{fig:feh_histo}
\end{figure*}

\section{Results} \label{sec:results}
\subsection{Age--metallicity relation}

\begin{figure}
\centering
\includegraphics[width=\columnwidth]{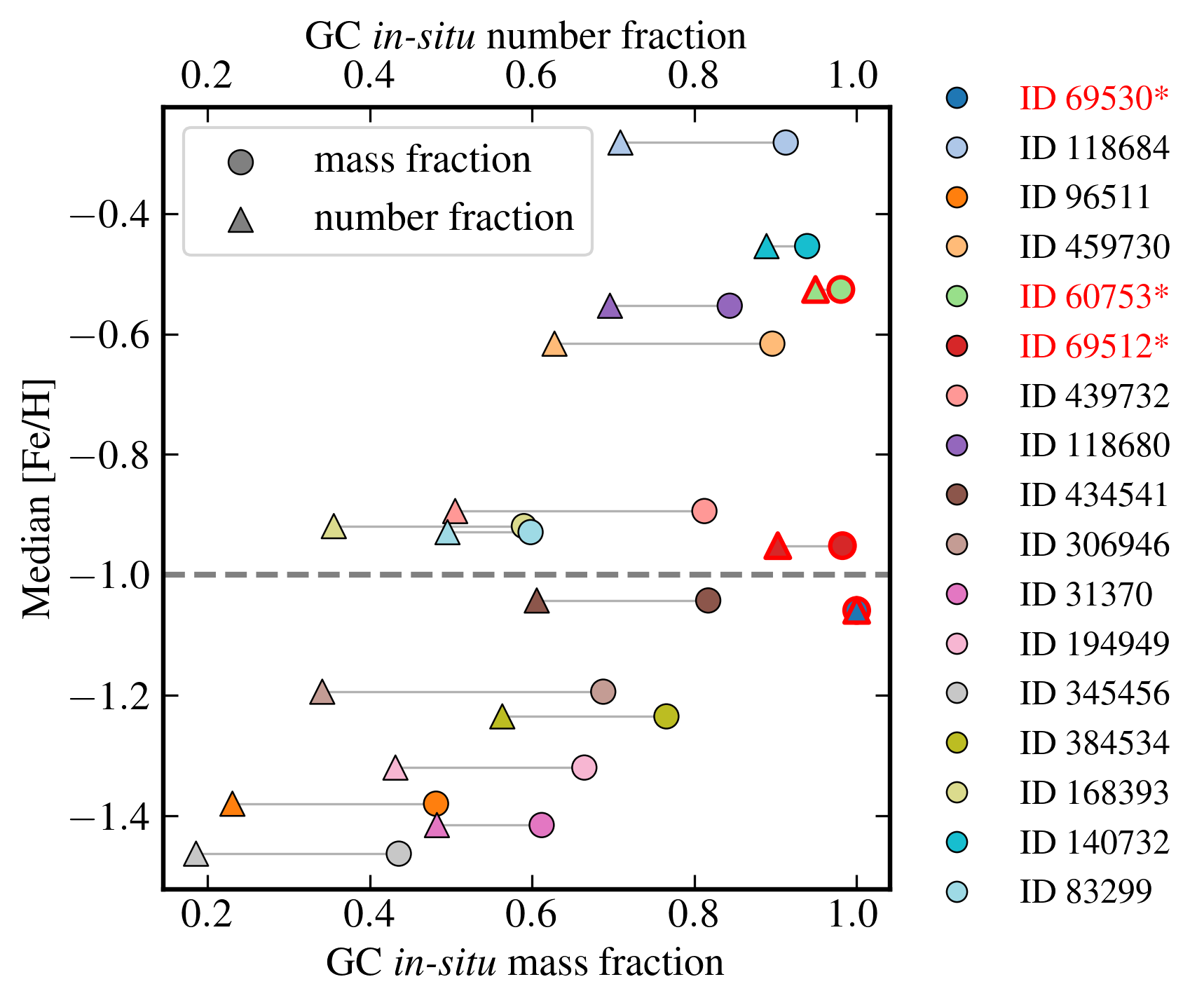}
\caption{The median metallicity ([Fe/H]) of the galaxies is plotted as a function of their \insitu GCs fraction. Each host galaxy is color-coded by its unique ID.}
\label{fig:feh_insitu}
\end{figure}

It is well known that the distribution of stellar populations with varying ages and metallicities in quiescent galaxies reveals the processes behind galaxy mass assembly across cosmic time. Observed trends in age and chemical enrichment constrain the dominant mechanisms of this growth (e.g.,\,\citealt{2013Cappellari,2014Gallazzi,2025Damjanov}). GCs with well-constrained ages and metallicities, offer a complementary perspective, where their age-metallicity relation reflects the chemical evolution of their host galaxies, making them powerful tracers of early star formation and enrichment histories (e.g.,\,\citealt{1978Searle, 2015Kruijssen,2022Reinacampos}). In the context of CMGs, this relation is thought to provide an additional constraint for identifying relic candidates, as their GC populations are expected to retain the chemical signatures of early assembly and reflect the absence of significant late-time accretion.

Fig.~\ref{fig:age-met_compacts} shows the age–metallicity distributions of GCs for each host galaxy, where clusters are color-coded by their origin: red for \insitu clusters formed within the main progenitor, and blue for \exsitu clusters accreted from external systems. A horizontal reference line at [Fe/H] = –1 is included to facilitate the comparison of metallicity trends across systems. The panels are ordered by decreasing host \insitu stellar mass fraction.

Despite differences in GC assembly histories, including variations in the number and metallicity of \exsitu accretions, the GC populations of all CMG hosts are predominantly old, with mean ages of $\sim$12\,Gyr, independent of their \insitu or \exsitu origin. In terms of metallicity, both components (\insitu and \textit{ex-situ}) can exhibit mean [Fe/H] values above –1, as seen in subhalo ID~83299, or below –1, as in IDs~69530 and~194949.

We quantified the metallicity spread among the 17 host galaxies shown in the last panel of Fig.~\ref{fig:age-met_compacts}. The dispersion ($\sigma$[Fe/H]) was computed as the standard deviation of the [Fe/H] distribution of the GCs within each host, providing a measure of the chemical inhomogeneity of the cluster system. The metallicity spread offers important clues about the assembly history of each galaxy, where a low dispersion suggests that the stellar content formed rapidly from chemically uniform gas, while a high dispersion implies that the system experienced multiple or extended star formation episodes, allowing progressive chemical enrichment of the interstellar medium. Thus, a narrow metallicity distribution points to a fast and homogeneous formation, whereas a broad distribution indicates a more prolonged and complex assembly history. As an example, host 384534 exhibits a distinct population of metal-rich GCs at $\sim6\,\rm{Gyr}$ that deviates from the main distribution. Given that this host experienced a significant stellar mass accretion ($\sim16\%$), these clusters were likely accreted during a significant merger event, originating from a companion galaxy that already hosted a metal-rich GC population. Consequently, the metallicity spread for 384534 is larger, consistent with a more prolonged and complex accretion history.

For our \insitu-dominated hosts (IDs 69530, 60753, 69512), we obtained a mean $\sigma$[Fe/H] = 0.357 and a median of 0.374, while the remaining hosts show higher mean and median dispersions of 0.562 and 0.498, respectively. When including the additional similar hosts (IDs 140732 and 118684), the mean $\sigma$[Fe/H] for the \insitu-dominated group becomes 0.427 (median = 0.374), compared to 0.567 (median = 0.498) for the others. These differences suggest that the GC \insitu–dominated hosts assembled more rapidly and in a more chemically homogeneous environment, consistent with a scenario of early and relatively short star formation history.

Another approach to analyse the spread of metallicities is shown Fig.~\ref{fig:feh_histo}. This figure ilustrates the individual GC metallicity distributions for each CMGs host. The GC populations spans a broad metallicity range, from metal-poor to metal-rich populations, with [Fe/H] values approximately between –2 and 0.

\begin{figure}
\centering
\includegraphics[width=\linewidth]{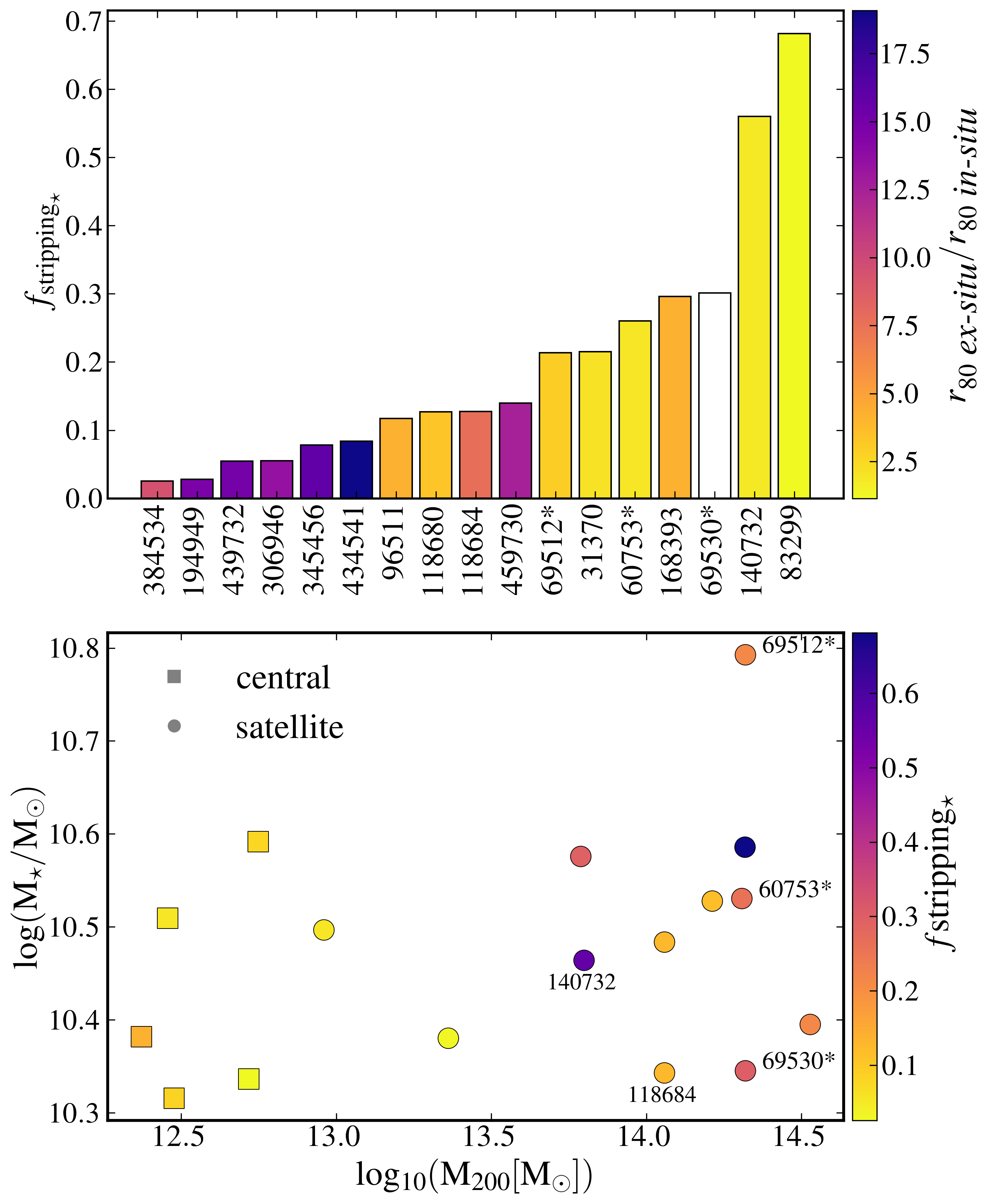}
\caption{Stellar stripping fraction and the environment of each CMGs. Upper panel: Stellar stripping fraction of each host color-coded by $r_{80~\textit{ex-situ}}/r_{80~\textit{in-situ}}$. This ratio quantifies the relative spatial extension of \exsitu and \insitu GCs, where $r_{80}$ denotes the radius enclosing 80\% of the respective population. The host 69530 is shown in white, as it contains no \exsitu globular clusters. Lower panel: Stellar mass as a function of $\rm{M}_{200}$ mass for each CMG color-coded by their stripping. circles and squares markers represent the satellite and central hosts, respectively.}
\label{fig:histogram_strip}
\end{figure}

It is particularly relevant to highlight the hosts that exhibit more extended star formation histories (see Fig.~\ref{fig:host_assembly_hist}), such as IDs 83299 and 140732. The GC metallicity distribution of host 83299 clearly reveals both \insitu and \exsitu GC components, whereas host 140732, despite having undergone a comparable stellar accretion fraction ($\approx20\%$), shows only a weak GC \exsitu signature. Both hosts experienced significant accretion and are also among those most affected by tidal stripping, which evidently alters the surviving GC population. Despite the stripping signatures present in their SFHs (Fig.~\ref{fig:host_assembly_hist}), it appears that only ID 140732 considerably lost its \exsitu component (becoming a highly \insitu GC dominated system), whereas ID 83299 retained a fraction of it. This is likely related to the fact that ID 83299 resides in a more massive $M_{200}$ halo compared to ID 140732 (lower panel of Fig.~\ref{fig:histogram_strip}).

In contrast, the relic-like hosts as 60753 and 69512 display similar distributions as 140732, characterized by a prominent, metal-rich \insitu component and a faint \exsitu contribution, as expected given our selection criteria. Unlike ID 140732, these systems did not experience significant mergers over time and therefore exhibit predominantly \insitu\ GC components due the absence of accretion.

Host 69530 did not undergo significant accretion events, yet its GC metallicity distribution is bimodal. In contrast, host 140732 is unimodal and dominated by \insitu GCs, but its evolutionary path includes important mergers and stripping events. These two cases contradict the usual understanding model of relic galaxies, which assumes that hosts with no accretion should contain exclusively \insitu GCs and therefore exhibit a unimodal metallicity distribution. This expectation matches the properties of 69530, yet in practice it aligns more closely with 140732.

This suggests that some observed relic candidates may display a wide range of \insitu metallicities, and that such spreads are not necessarily driven solely by accretion, although such cases require further investigation. Concerning host galaxies such as 140732, they represent the `stripped CMGs', which are galaxies that are not strictly frozen in time, yet still reproduce the mass, size, and GC distribution expected for genuine relics. In the next section, we introduce a method to distinguish stripped from non-stripped galaxies.

To summarize the metallicity trends, Fig.~\ref{fig:feh_insitu} presents the median [Fe/H] values for all CMG hosts depending on GC \insitu mass and number fraction for reference. While \citet{2014Harris} do not explicitly compare number and mass‑fractions of \insitu GCs, their finding of a nearly constant GC system characteristic mass (from the luminosity function) implies that the number of GCs is a reliable proxy for the total GC mass in massive galaxies. In our analysis, the small difference between number and mass fractions of \insitu GCs suggests that accretion from low-mass satellites is minor for our CMGs. We note, however, that for consistency we apply a cut selecting only GCs with masses $>10^{5} \rm{M_\odot}$ to avoid spurious low-mass objects, which could introduce a bias in the number fraction.

As a result, using [Fe/H]$=-1$ as an arbitrary reference threshold to distinguish between the metal-rich and metal-poor regimes, we find that 52.9\% of the hosts have median [Fe/H] above this value, while 47.1\% fall below it. The three relic-like hosts are, on average, more metal-rich than the remaining fourteen. The mean [Fe/H] was computed by averaging the individual [Fe/H] values of the hosts, 60753, 69512 and 69530 have a mean [Fe/H] of $\sim-0.896$, while the remaining fourteen galaxies have a mean of $\sim-1.062$, with a difference of 0.166 dex between both groups. When we consider also the 140732 and 118684 among the relic-like hosts, we obtain a mean [Fe/H] of $\sim-0.763$ for this group, and $\sim-1.145$ for the remaining hosts, adding a difference of 0.383 dex between the groups. As a result, the selected relic-like hosts are, on average, more metal-rich and predominantly with unimodal distribution, as for NGC\,1277. 

We applied an empirical color-metallicity relation to Fig.~\ref{fig:age-met_compacts} and \ref{fig:feh_histo} (see Appendix~\ref{appendixA}, in Fig.~\ref{fig:A1} and \ref{fig:A2}) as an exploratory test, intended primarily to assess whether the simulated GC populations show agreement with observational trends reported for relic galaxies (e.g., \citealt{2018Beasley}). We adopt the \citet{2010ABlakeslee} $g-z$ calibration, which provides a reliable empirical mapping for old stellar systems, restricting the input metallicities to $-2.5 < [\rm{Fe/H}] < 0.2$ and the resulting colors to the calibrated range $0.8 < (g-z) < 1.4$. Values outside these domains were discarded to avoid extrapolations beyond the empirically supported regime. Due to colors being far more accessible observationally than metallicities for GC systems in distant or compact hosts, these conversions also offer a practical way to generate testable predictions for future comparisons with real GC samples.

\begin{figure}
\centering
\includegraphics[width=\columnwidth]{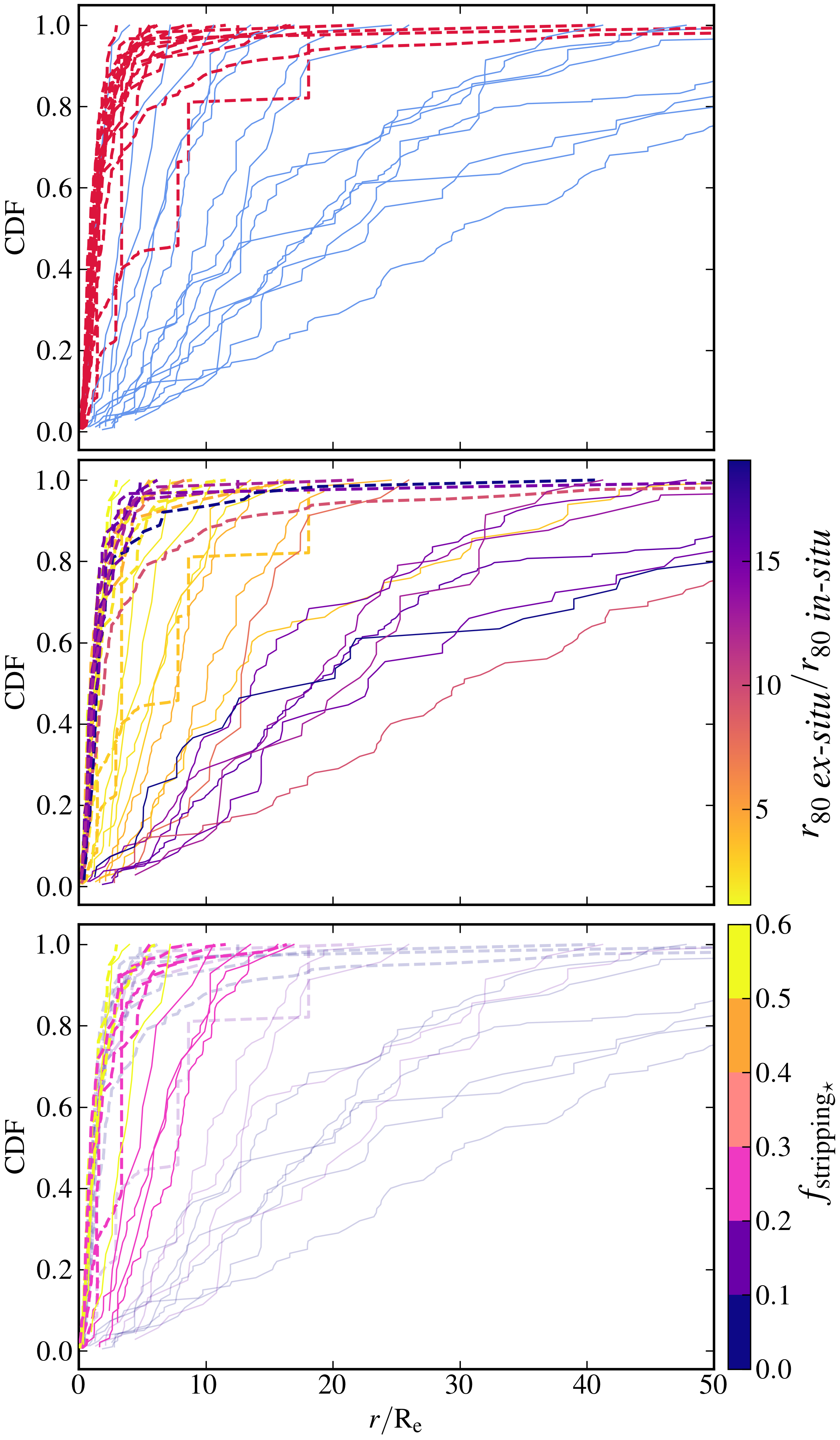}
\caption{Combined cumulative distribution function (CDF) for all the hosts. In the first row, the components are color-coded by \insitu and \exsitu GCs in red (dashed) and blue (solid) lines, respectively. Second row shows the CDF color-coded by the $r_{80~\textit{ex-situ}}/r_{80~\textit{in-situ}}$ ratio, where lower values represent small galactocentric distances for the \exsitu component. The third row shows the CDF color-coded by the stellar stripping fraction, with a highlight for cases where the stripping is pronounced $f_{\rm{stripping}\star}>0.2$. In all the rows the
the dashed lines represent the \insitu GC component of each host.}
\label{fig:CDF2}
\end{figure}
\begin{figure*}
\centering
\includegraphics[width=\linewidth]{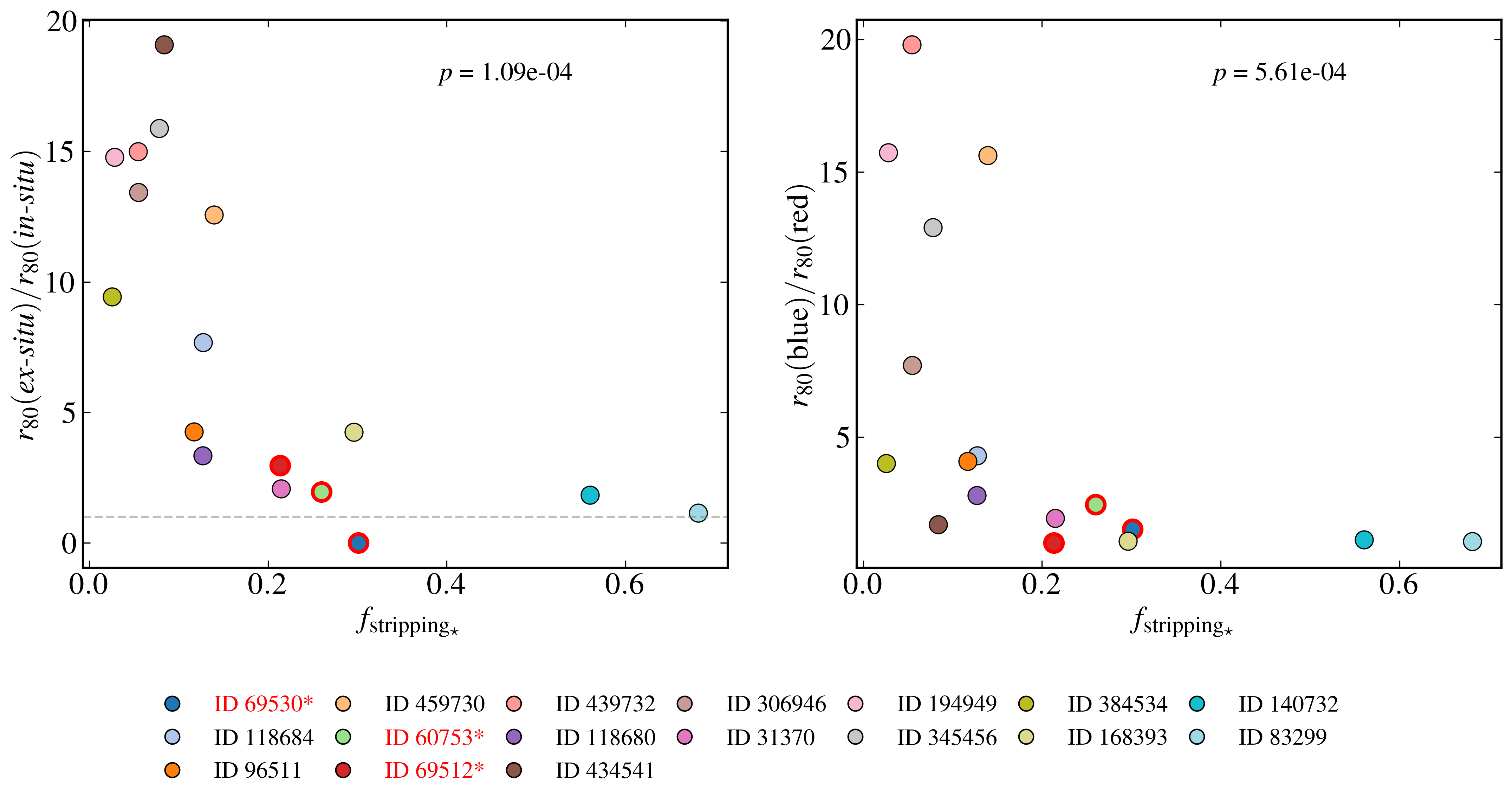}
\caption{Relative galactocentric distance for \insitu and \exsitu component quantified by $r_{80}$ ratio depending on host stellar stripping fraction. The gray dashed line indicates where both, \insitu and \exsitu population have comparable galactocentric extents.} 
\label{fig:R80ratio_strip}
\end{figure*}
\subsection{Stellar stripping and spatial distributions}
\label{stripping}

The previous discussion on the chemical evolution of GCs highlights the imprints on the SFHs of their hosts. Yet, these signatures can be further modified by subsequent interactions and mass loss. Stellar tidal stripping, in particular, can remove part of the stellar and GCs components, reshaping the observed structural and chemical properties of the remnant galaxy (i.e.,~\citealt{2015Bastian,2019Li}). The stellar stripping is entangled with the dynamics and spatial distribution of GCs in their hosts. The GCs relative positions can provide clues to distinguish between compact galaxies showing stripping signatures and those that do not. This approach could be particularly valuable for studying relics that lost their envelope versus those that did not accrete additional material. Regarding these cases, \citet{2025Zhu} investigated compact ETGs and found stellar dynamical signatures among stripped cases.

Here, we define the host stellar stripping fraction as: $1 - (\rm{M_{\star (\textit{z}=0)}} / \rm{M_{\star (\mathrm{peak})}})$. Given that stellar stripping affects the distribution of stars in the galactic outskirts, we also assume it impacts the GC distribution. Fig.~\ref{fig:histogram_strip} (upper panel) quantifies the stellar tidal stripping fraction for all host CMGs, color-coded by the ratio between $r_{80}$ (radius enclosing 80\% of GCs) for the \insitu and \exsitu GC population. This ratio quantifies how much the \exsitu population is extended in comparison with the \insitu population. Subhalo 69530, although showing a stripping fraction of $\sim$30\%, does not have an \exsitu component and is therefore displayed without color coding in the upper panel. The lower panel of Fig.~\ref{fig:histogram_strip} shows the $M_\star-M_{200}$ relation for each host and their classification into central and satellite within their halos for reference. One expects that for highly stripped cases (i.e., 83299, 140732) the galactocentric distance of the \exsitu population is closer to the \insitu population.

To better constrain and to quantify this trend, Fig.\,\ref{fig:CDF2} shows the normalized galactocentric distance of the \insitu and \exsitu population of GCs for all hosts. We find that the \exsitu component is significantly more extended than the \insitu population (first row, Fig.\,\ref{fig:CDF2}), with a median $r_{80}$ ratio of $6.0$. On average, \exsitu GCs are $8.1\times$ more extended than their \insitu counterparts (second row). However, galaxies with high stellar stripping fractions ($f_{\rm{stripping_\star}} > 0.2$, highlighted hosts in third row) show much smaller $r_{80}$ ratios (median of $2.0$, mean of $2.4\times$), indicating that \exsitu GCs can be found at relatively small galactocentric distances. This compact distribution of the \exsitu population may therefore represent a signature or proxy of stellar stripping processes in a given host galaxy. 

Fig.\,\ref{fig:R80ratio_strip} shows the dependence of the $r_{80}$ ratio with the host stellar stripping fraction. The dashed line at ratio = 1 marks the case where both GC components have comparable spatial extents. Hosts above this line exhibit more extended \exsitu GC distributions, whereas those near or below it indicate comparable or more compact \exsitu components. A trend is observed in which galaxies with higher stripping fractions tend to have ratios close to this line, consistent with the removal of the outer \exsitu population. The Spearman rank test confirms a strong negative correlation ($\rho =-0.82$), with $p-\rm{value} = 1.1\times10^{-4}$, indicating that increasing stripping leads to progressively less extended \exsitu\ distributions. In turn, galaxies with lower stellar stripping fractions display larger ratios, suggesting more extended \exsitu components and possibly less dynamically evolved halos.

\section{Discussion and summary}
\label{sec:discussion}

Our analysis of 17 compact massive hosts from TNG100 reveals a diverse set of assembly histories and their associated GC populations. By combining information from stellar mass growth, effective radius evolution, GC origin (\insitu \textit{vs.} \exsitu), metallicity, and spatial distributions, we can draw a consistent picture of early galaxy assembly, late-time accretion, signatures of stellar stripping and ther correlation with the GCs populations.  

\subsection{GCs as tracers of host assembly}

The comparison between the \insitu fractions of hosts and their GCs demonstrates that the GC mass fraction is a more robust tracer of the host's assembly history than the GC number fraction, showing a lower average deviation from the identity line parameter shown in Fig.~\ref{fig:similarity2}. This result suggests that the mass-weighted contribution of GCs more reasonably reflects the host's accretion history, likely because massive clusters are less susceptible to disruption and tidal effects \citep{2019Choski,Caso2024}

Three CMGs (IDs 60753, 69512, 69530) consistently emerge as extreme cases, with high \insitu fractions in both the host and GC populations, representing the most promising relic galaxy candidates. Two additional hosts (140732 and 118684) initially seemed to be promising candidates. However, 140732 turned out to be a stripped relic, since its SFH is extended, and it has lost most of its content through tidal interactions. In turn, ID 118684 host appears to have accreted only a small number of satellites (less than 5\%), yet the accreted GCs are very massive, producing a prominent \exsitu blue GC component in Fig.~\ref{fig:feh_histo}. This behaviour is reminiscent of the few massive, blue GCs in the Fornax dSph \citep{2012Larsen}, and more recently, of the `cosmic gems' described by \citet{2025Vanzella}. These highlighted cases provide a benchmark for understanding the implication of different SFHs on the GCs population and distribution.

\subsection{Chemical signatures and age--metallicity relations}

The age–metallicity distributions of the GCs reveal that all CMG hosts host predominantly old populations ($\sim$12\,Gyr), consistent with rapid early assembly. The metallicity spread among the \insitu--dominated relic candidates is narrower ($\sigma$[Fe/H] $\sim 0.36-0.43$) than in the other CMGs ($\sigma$[Fe/H] $\sim 0.50-0.57$). This supports the interpretation that the \insitu--dominated hosts formed rapidly from chemically homogeneous gas, whereas the remaining systems possibly experienced more prolonged or multiple star formation episodes.

In terms of median metallicities, the three relic-like hosts are systematically more metal-rich ([Fe/H]$\sim -0.896$ for the three most extreme cases, or $\sim -0.763$ including 140732 and 118684) than the remaining galaxies ([Fe/H]$\sim -1.062$ or $\sim -1.145$), supporting the scenario where relic galaxies tend to have more metal-rich GC populations \citep{2018Beasley,2021Kanglee}.

\subsection{Stellar stripping and GC spatial distributions}

Stellar tidal stripping plays a crucial role in shaping the observed properties of both stars and GCs. We find an anti-correlation between the host stellar stripping fraction and the relative extension of the \exsitu GC population ($r_{80,\,\rm ex-situ}/r_{80,\,\rm in-situ}$), with a Spearman rank correlation of $\rho=-0.82$. Highly stripped systems (e.g., 83299, 140732) exhibit \exsitu GCs that are relatively compact, approaching the spatial distribution of \insitu clusters, whereas minimally stripped hosts retain extended \exsitu populations. This result highlights the potential of GC spatial distributions as a diagnostic of past tidal stripping events.

\subsection{Implications for relic galaxy identification}

The combination of high \insitu fractions, narrow metallicity spreads, and compact GC distributions provides a robust set of criteria for identifying relic galaxies. The three most extreme systems (60753, 69512, 69530) exhibit consistently high \insitu fractions, relatively metal-rich and homogeneous \insitu populations, and minimal \exsitu contamination, making them ideal candidates for local analogs relic galaxies. Among these three, ID 69530 shows the most complex GC assembly. It has a broad metallicity spread even though it experienced no significant mergers (no \exsitu GCs).

Two additional hosts, 140732 and 118684, are also quite interesting. ID 140732 is a stripped relic, but it still displays strong relic-like features similar to the three main systems. ID 118684, in contrast, had only a small amount of accretion, yet it accreted very massive GCs. This produced a bimodal and broader metallicity distribution, with a significant metal-poor component. Despite this, ID 118684 would be classified as a good relic candidate in terms of their other properties as mass, size, and its low accretion fraction ($<5\%$).

Overall, our work reinforces the value of GC populations as complementary tracers of galaxy assembly. By connecting stellar assembly histories, GC chemical properties, and spatial distributions, we are able to reconstruct the early formation pathways of CMGs and to identify robust relic galaxy candidates. In a follow-up work we plan to delve into the GC dynamics signatures of the CMGs analysed in the present work. Complementarily, future observational follow-ups of these systems, particularly in terms of GC kinematics and high-resolution metallicity mapping, are of utmost importance to provide constraints on their formation histories and the role of environmental processes such as stripping in shaping their present-day properties.

\section*{acknowledgments} 
\label{sec:acknows}
 We thank the referee for the suggestions, which have helped make the manuscript more robust. MTM acknowledges the Brazilian agencies: Conselho Nacional de Desenvolvimento Científico e Tecnológico (CNPq) through grant 140900/2021-7, Coordenação de Aperfeiçoamento de Pessoal de Nível Superior (CAPES) through 88887.936624/2024-00 and e Fundação de Amparo à Pesquisa do Estado de São Paulo (FAPESP) through grant 2025/09192-7. ACS acknowledges funding from CNPq (grants 314301/2021-6 and 445231/2024-6) and the Rio Grande do Sul State Research Foundation (FAPERGS, 24/2551-0001548-5). CF ackwoledges funding from CNPq and FAPERGS through grants CNPq-315421/2023-1 and FAPERGS-21/2551-0002025-3. AFM acknowledges support from RYC2021-031099-I and PID2024-162088NB-I00 of MICIN/AEI/10.13039/501100011033/ UE NextGenera-
tionEU/PRTR. JPC acknowledges funding from Consejo Nacional de Investigaciones
Cient\'{\i}ficas y T\'ecnicas de la Rep\'ublica Argentina (PIP 112-2020010-1567)
and Universidad Nacional de La Plata (PPID 11/G183 2023-2026, Argentina).

\clearpage

\bibliography{ref}
\bibliographystyle{aasjournal}
\clearpage
\appendix
\label{appendixA}
\clearpage

\begin{table*}
\centering
\caption{Global properties of the GC systems for each host galaxy. The first column lists the host IDs. Columns~2 and~3 show the \insitu and \exsitu GC fractions computed by number ($N_{\mathrm{GC}}$), while columns~4 and~5 present the same fractions computed by mass. Columns~6–8 provide the total number of GCs and the corresponding numbers of \insitu\ and \exsitu\ clusters. The last column lists the logarithm of the total GC mass ($\log M_{\mathrm{GC}}/M_{\odot}$).}

\begin{tabular}{lcccccccc}
\hline\hline
ID & $\rm{N}f_{\textit{\insitu}}$ & $\rm{N}f_{\textit{\exsitu}}$ &  $\rm{M}f_{\textit{\insitu}}$ & $\rm{M}f_{\textit{\exsitu}}$ & $N_{tot}$ & $\rm{N}_{\textit{\insitu}}$  &  $\rm{N}_{\textit{\exsitu}}$  &  $\log\,(\rm{M_{GC}}/\rm{M}_{\odot})$ \\
\hline\hline
31370 & 0.483 & 0.517 & 0.612 & 0.388 & 116 & 56 & 60 & 7.946 \\
60753 & 0.949 & 0.051 & 0.980 & 0.020 & 197 & 187 & 10 & 8.319 \\
69512 & 0.903 & 0.097 & 0.982 & 0.018 & 144 & 130 & 14 & 8.196 \\
69530 & 1 & 0 & 1 & 0 & 45 & 45 & 0 & 7.596 \\
83299 & 0.496 & 0.504 & 0.598 & 0.402 & 119 & 59 & 60 & 8.065 \\
96511 & 0.231 & 0.769 & 0.482 & 0.518 & 91 & 21 & 70 & 7.613 \\
118680 & 0.696 & 0.304 & 0.844 & 0.156 & 273 & 190 & 83 & 8.184 \\
118684 & 0.709 & 0.291 & 0.913 & 0.087 & 79 & 56 & 23 & 7.735 \\
140732 & 0.889 & 0.111 & 0.939 & 0.061 & 189 & 168 & 21 & 8.287 \\
168393 & 0.356 & 0.644 & 0.590 & 0.410 & 149 & 53 & 96 & 8.057 \\
194949 & 0.432 & 0.568 & 0.664 & 0.336 & 139 & 60 & 79 & 7.885 \\
306946 & 0.341 & 0.659 & 0.688 & 0.312 & 123 & 42 & 81 & 7.826 \\
345456 & 0.186 & 0.814 & 0.436 & 0.564 & 242 & 45 & 197 & 8.048 \\
384534 & 0.563 & 0.437 & 0.766 & 0.234 & 229 & 129 & 100 & 8.070 \\
434541 & 0.606 & 0.394 & 0.817 & 0.183 & 104 & 63 & 41 & 7.787 \\
439732 & 0.505 & 0.495 & 0.812 & 0.188 & 95 & 48 & 47 & 7.750 \\
459730 & 0.628 & 0.372 & 0.896 & 0.104 & 94 & 59 & 35 & 7.795 \\
\hline\hline
\end{tabular}
\label{tab:gc_fractions}
\end{table*}

\begin{figure}
\centering
\includegraphics[width=\columnwidth]{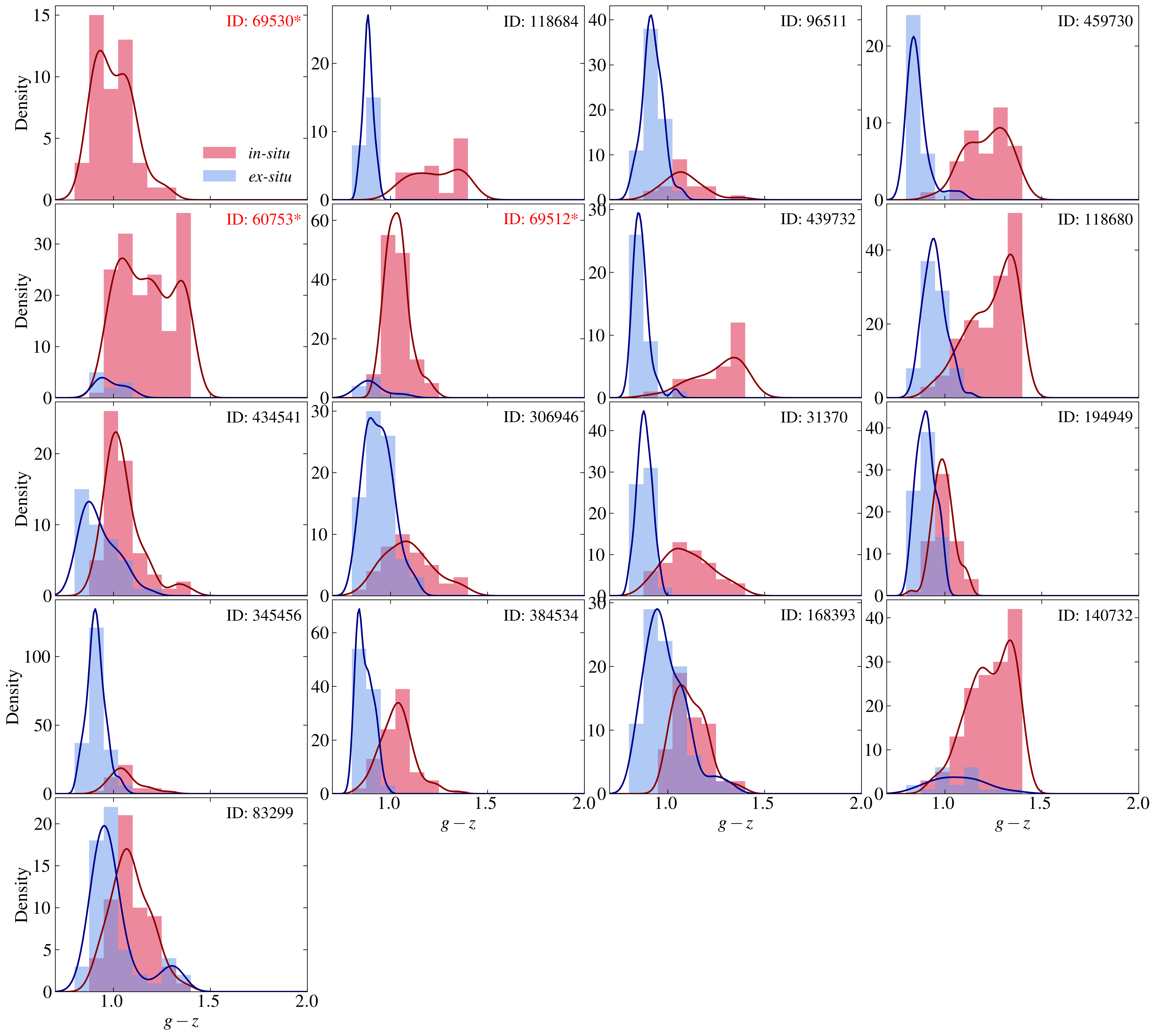}
\caption{Histograms of GC colors ($g-z$) for each host galaxy in the sample, following the same scheme as in Fig.~\ref{fig:feh_histo}. The distributions are color-coded in red and blue to represent in-situ and ex-situ populations, respectively.}
\label{fig:A1}
\end{figure}

\begin{figure}
\centering
\includegraphics[width=\columnwidth]{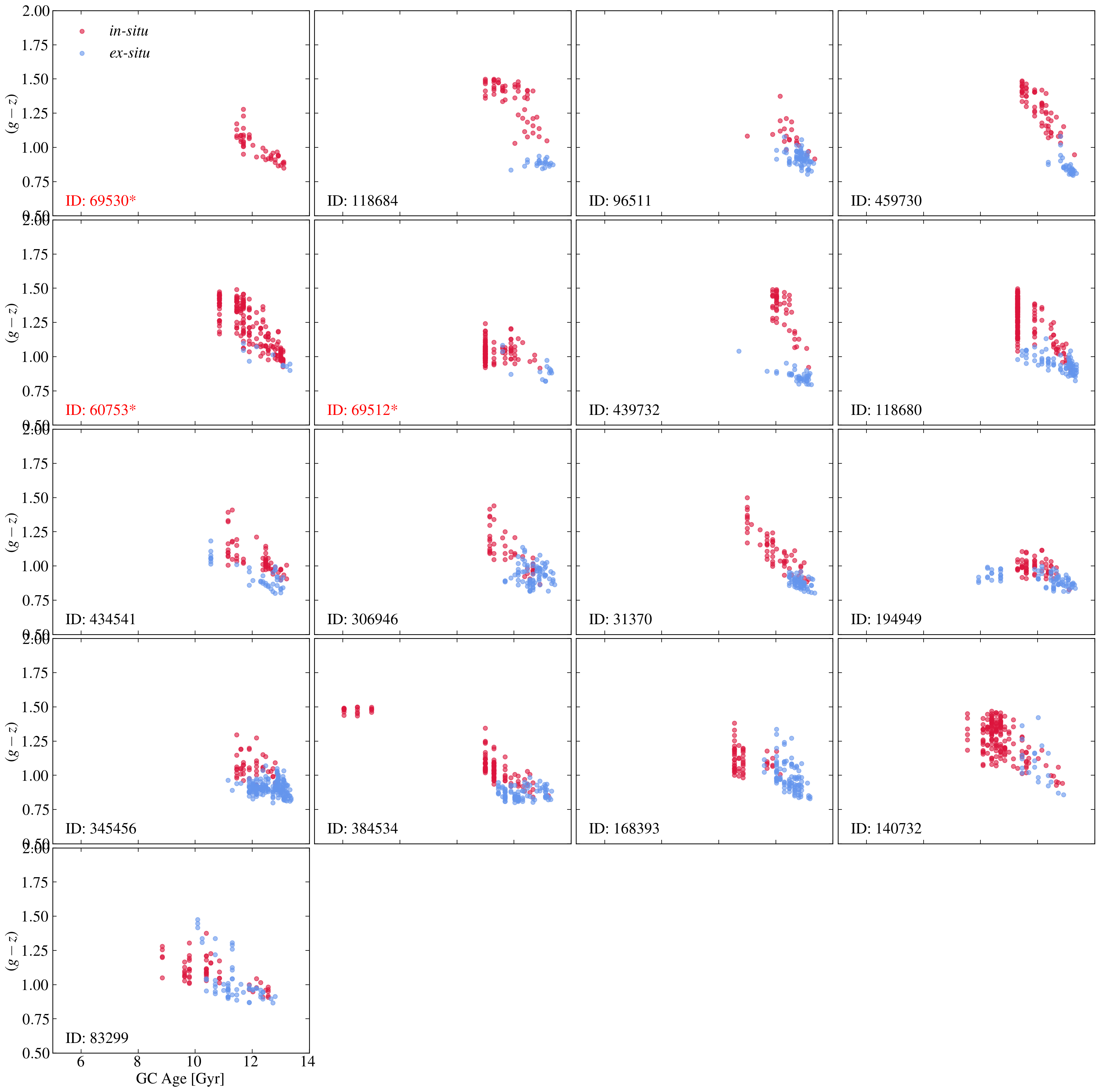}
\caption{$g-z$ color distribution over ages for all GCs in each host galaxy. GCs in-situ are displayed in red, and ex-situ GCs are displayed in blue in all the frames. The Most GC in-situ -dominated hosts are highlighted with ‘*’ symbol.}
\label{fig:A2}
\end{figure}

\begin{figure}
\centering
\includegraphics[width=\columnwidth]{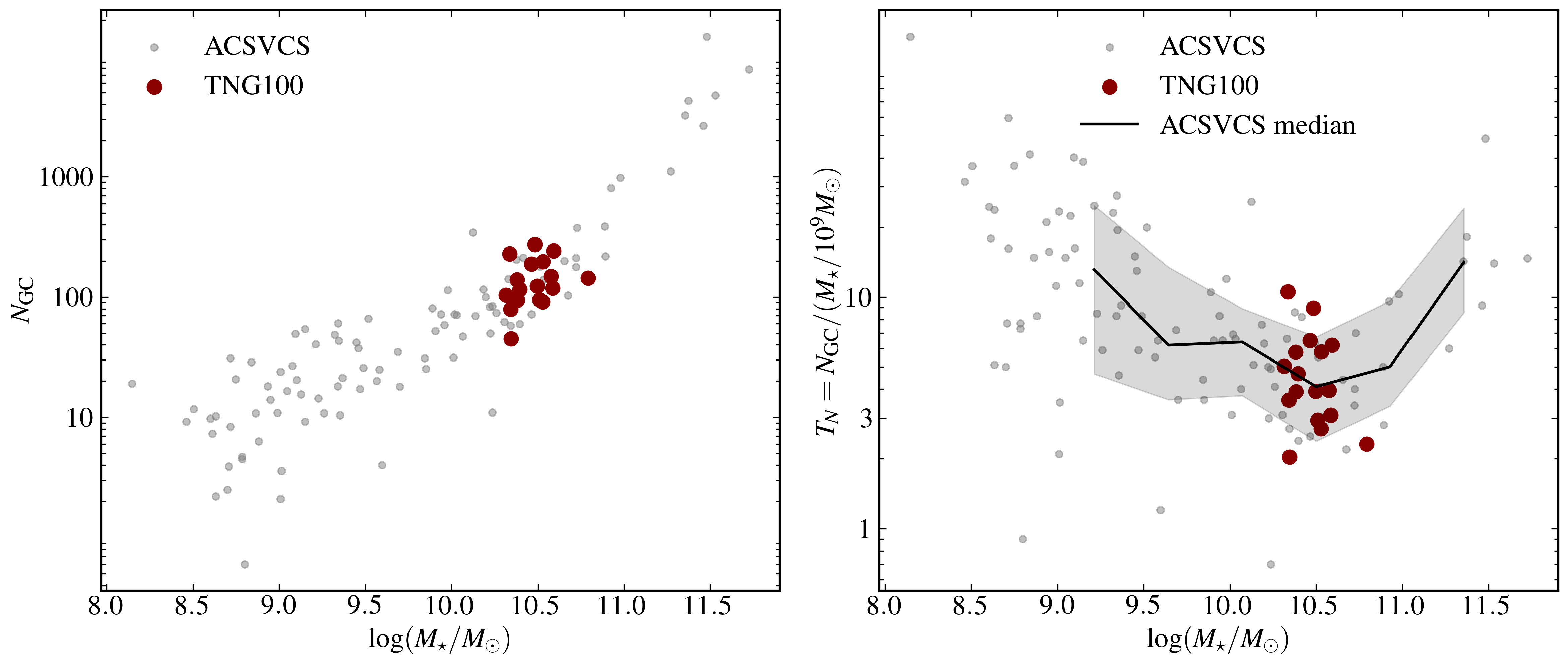}
\caption{$N_{\rm{GC}}$ and $T_N$ data from ACSVCS among the host sample from TNG100. Grey points represent individual ACSVCS galaxies. The black line shows the median for $T_N$ in bins of stellar mass, while the shaded region indicates the 16th–84th percentile range of the observed distribution on the right panel. The host CMGs (red points) lie within this observational range and follow the median relation, indicating that the number of GCs in the simulations is consistent with that observed in galaxies of similar stellar mass.
}
\label{fig:A3}
\end{figure}





\end{document}